\documentclass[aps,prb,twocolumn,showpacs,groupedaddress]{revtex4}

\usepackage{graphicx}

\newcommand{\etal}{{\it et~al.}}

\begin{document}

\preprint{}

\title{Reversible Metal-Insulator Transition in LaAlO$_3$ Thin Films Mediated by Intragap Defects: An Alternative Mechanism for Resistive Switching}

\author{Z. Q. Liu$^{1,2}$}

\author{D. P. Leusink$^{1,2}$}

\author{W. M. L\"{u}$^{1,3}$}

\author{X. Wang$^{1,2}$}

\author{X. P. Yang$^{4}$}

\author{K. Gopinadhan$^{1,3}$}

\author{Y. T. Lin$^{1,2}$}

\author{A. Annadi$^{1,2}$}

\author{Y. L. Zhao$^{1,2}$}

\author{A. Roy Barman$^{1,2}$}

\author{S. Dhar$^{1,3}$}

\author{Y. P. Feng$^{1,2}$}

\author{H. B. Su$^{4}$}

\author{G. Xiong$^{1,3}$}

\author{T. Venkatesan$^{1,2,3}$}

\author{Ariando$^{1,2}$}

\altaffiliation[Email: ]{ariando@nus.edu.sg}

\affiliation{$^1$NUSNNI-Nanocore, $^2$Department of Physics, $^3$Department of Electrical and Computer Engineering, National University of Singapore, Singapore}

\affiliation{$^4$Division of Materials Science, Nanyang Technological University, Singapore}

\date{\today}

\begin{abstract}
We report the electric-field-induced reversible metal-insulator transition (MIT) of the insulating LaAlO$_3$ thin films observed in metal/LaAlO$_3$/Nb-SrTiO$_3$ heterostructures. The switching voltage depends strongly on the thickness of the LaAlO$_3$ thin film which indicates that a minimum thickness is required for the MIT. A constant opposing voltage is required to deplete the charges from the defect states. Our experimental results exclude the possibility of diffusion of the metal electrodes or oxygen vacancies into the LaAlO$_3$ layer. Instead, the phenomenon is attributed to the formation of a quasi-conduction band (QCB) in the defect states of LaAlO$_3$ that forms a continuum state with the conduction band of the Nb-SrTiO$_3$. Once this continuum (metallic) state is formed, the state remains stable even when the voltage bias is turned off. The thickness dependent reverse switch-on voltage and the constant forward switch-off voltage are consistent with our model. The viewpoint proposed here can provide an alternative mechanism for resistive switching in complex oxides.
\end{abstract}

\pacs{73.40.Rw, 73.50.Gr, 73.20.Hb}


\maketitle

The electronic state of complex oxides can be changed by chemical doping, temperature, external pressure [1-3], magnetic field [4], electric field [5] or light [6]. In particular, the electric-field-induced MIT has attracted a lot of attention because of its intriguing physical mechanisms [7,8] and potential for device applications.  With the recent excitement in resistive switching [9-11] in typical metal/insulator/metal structures [12,13], electric-field-induced MIT has been revisited as a possible mechanism [14-16]. A number of different mechanisms [17-20] have been previously demonstrated for resistive switching, such as electric-field-induced oxygen vacancy migration resulting in the formation of conducting filaments [17,18], and reversible metal migration from electrodes [19]. Such a variety of field-induced phenomena arise from the complex defects that are present in these oxides. Cationic and anionic defects can form trap states within the bandgap of even wide bandgap oxides, drastically affecting their insulating properties [21]. In this paper, we show reversible MIT caused by the electrical population of defect levels in the bandgap of oxide insulators, which in turn can play a crucial role in determining the insulating nature of the material. The very large bandgap of LaAlO$_3$ (LAO) of $\sim 5.6$ eV [22], and the possibility of fabricating high crystalline quality thin films make it an ideal material for investigation of the unusual insulating properties of complex oxides.

\begin{figure}
\includegraphics[width=3.3in]{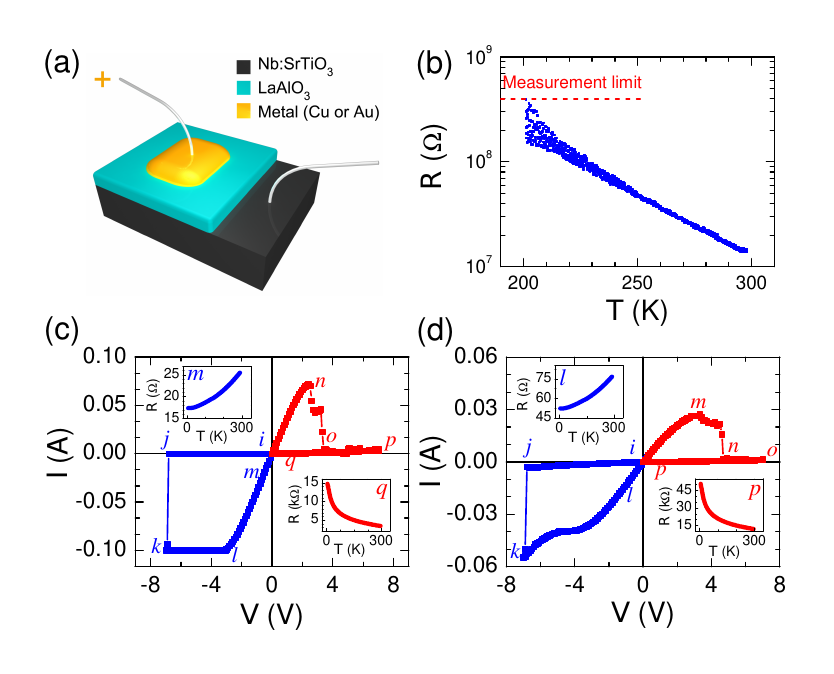}
\caption{\label{fig1}(a) Sample configuration and measurement geometry. (b) $R-T$ curve of the initial resistance state for Cu/LAO ($\sim$150 nm)/NSTO structure on a semi-logarithmic scale. (c) $I-V$ measurements by scanning voltage along $0\rightarrow -7$ V $\rightarrow$ 0 $\rightarrow$ 7 V $\rightarrow$ 0 and $R-T$ curves of different resistance states. The inset $m$ and $q$ are the $R-T$ curves after scanning the voltage through points $m$ and $q$ to zero, respectively. The horizontal data between points $k$ and $l$ are due to the compliance current in action. (d) $I-V$ curves after the cycle shown in (c) and $R-T$ curves of different resistance states. The inset $l$ and $p$ are the $R-T$ curves after scanning the voltage through points $l$ and $p$ to zero, respectively.}
\end{figure}

Typical metal/insulator/metal structures (Fig. 1a) were prepared by first depositing LAO films using a pulsed laser deposition technique from a single crystal LAO target onto (001)-oriented 0.5 wt\% Nb-doped SrTiO$_3$ (NSTO) single-crystal substrates. The LAO films of various thicknesses ($25-150$ nm) were grown at 800$^\circ$C under an O$_2$ partial pressure of $1\times10^{-2}$ Torr. During deposition, the fluence of the laser energy was 1.3 J/cm$^2$ and a shadow mask was used to cover a small part of NSTO to be used as the bottom electrode. After deposition, the sample was annealed for 1 hour at 600$^\circ$C in ambient O$_2$. Subsequently, the samples were cooled to room temperature at a very slow ramping rate of 5$^\circ$C/min to avoid cracking of the films, which can occur because of the mismatch between the thermal expansion coefficients of LAO and NSTO. Finally, $1\times 2$ mm$^2$ metal top electrodes (Cu or Au) were deposited on top of the LAO film by pulsed laser deposition through a stencil mask. The temperature dependence of the sample resistance ($R-T$) was measured by applying a voltage of 0.05 V between the top metal electrode and bottom electrode (Fig. 1a). At this small voltage, the existing resistance state is stable over time. Using the same two-terminal geometry, current-voltage ($I-V$) characteristics were measured by scanning the applied voltage and reading the current with a compliance current of 0.1 A. The sign of voltage corresponds to the sign of the voltage applied to the metal top electrode.

Figure 1 shows that the electronic phase of the Cu/LAO (150 nm)/NSTO structure can be reversibly changed by the applied voltages. In the following, we will describe the switching sequence starting with the negative voltage. As illustrated in Fig. 1(b), the temperature dependence of the $R-T$ curve of the initial state indicates typical insulating behavior that corresponds to the initial $I-V$ curve of the system as seen from $i$ to $j$ in Fig. 1(c). At low voltages, the resistance is very high but when the voltage reaches $-6.8$ V, a sharp jump in the current is seen. Because of the imposed compliance current of 0.1 A, which is used to prevent sample damage [12,19], the current saturates at 0.1 A. As the voltage is scanned back to 0 V, a linear $I-V$ characteristic is seen, which corresponds to a stable metallic state shown in the inset $m$ of Fig. 1(c). Hence, upon the dramatic switching of the resistive state triggered at -6.8 V, the resistance changes from $\sim 14$ M$\Omega$ to $\sim 25$ $\Omega$ at room temperature, which is concomitant with a phase transition from the insulating to the metallic phase. The metallic state persists until the positive voltage scan reaches about 2.4 V; at that level, the structure transitions into a high-resistance ($n-o-p-q$) and non-metallic state as indicated in the inset $q$ of Fig. 1(c). The reversible phase transition is reproducible, as shown in the second $I-V$ cycle in Fig. 1(d). These behaviors are totally unexpected for a wide bandgap insulator such as LAO.

The anomalous insulating behavior of LAO cannot be explained by artefacts such as anodization or redox of the active metal electrode [19]. In those cases, anodic dissolution of the metal electrode is possible only if adequate positive voltage is applied to an active metal electrode, and the resulting cations can be driven by the strong electric field into the insulating film where they form metallic filaments. To further eliminate this possibility, structures with inactive Au as the top electrode material were prepared and analyzed; these structures showed no observable difference in their resistive switching behavior compared to those that had Cu as the top electrode (supporting material). This strongly suggests that the reversible phase transition cannot be caused by the diffusion of metal electrodes.

\begin{figure}
\includegraphics[width=3.3in]{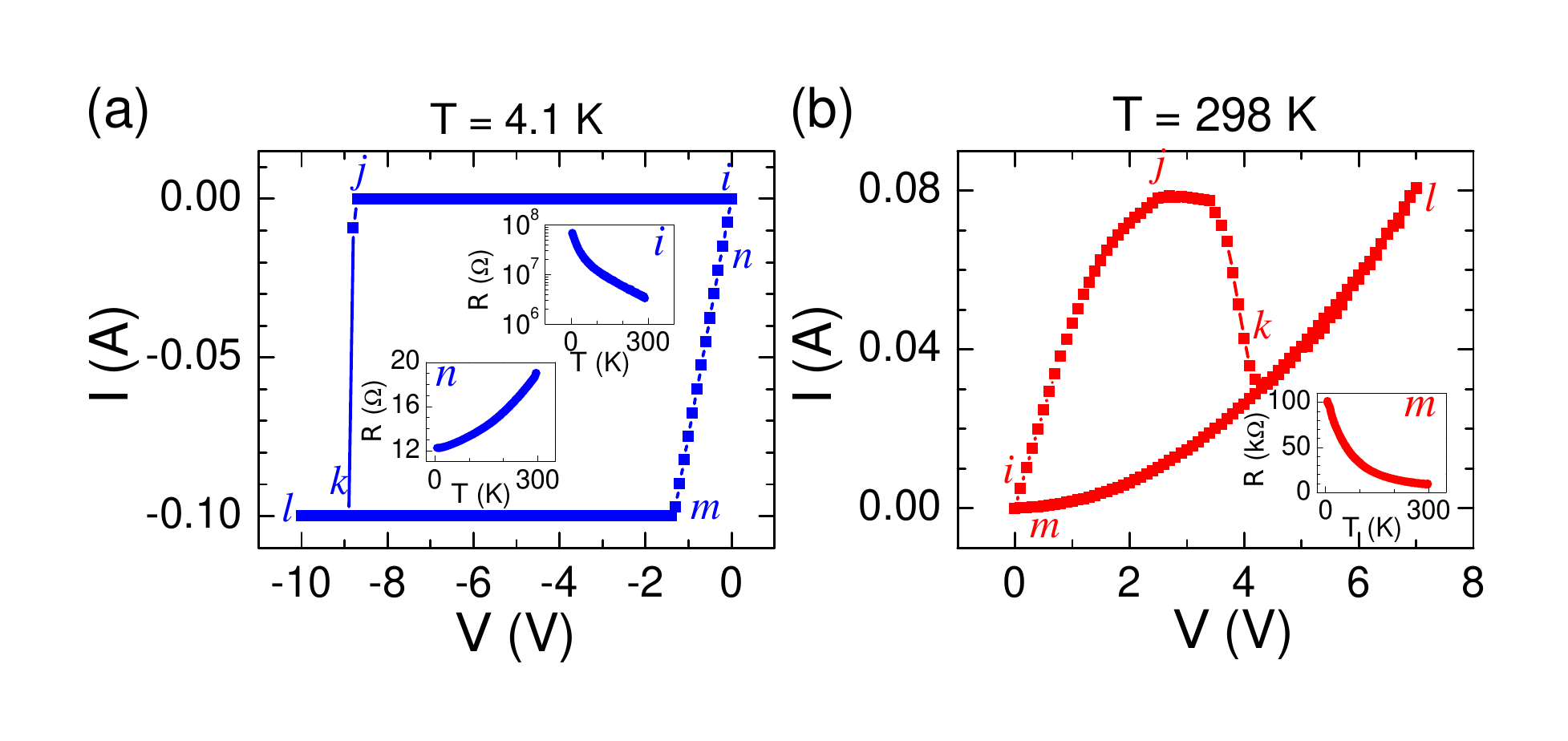}
\caption{\label{fig2}(a) Negative switching of Au/LAO ($\sim$150 nm)/NSTO at 4.1 K. The inset $i$ and $n$ are $R-T$ curves of the initial state and after negatively scanning voltage back to 0 ({\it{i.e.}}, after 0 $\rightarrow$ $-10$ V $\rightarrow$ 0 at 4.1 K), respectively. The current values in the $k-l-m$ sequence are confined by the compliance current. (b) Positive scan after (a) at 298 K by $0 \rightarrow  7$ V $\rightarrow  0$. The inset $m$ is $R-T$ curve after positive scan.}
\end{figure}

Another possible mechanism for this anomalous insulating phenomenon is the migration of oxygen vacancies. If such a mechanism is occurring, the applied electric field could only change the distribution of the positively charged oxygen vacancies [23]. Therefore, no insulating phase would appear, even at higher resistance states. Our case is also significantly distinct from the type of resistive switching that originates from the electro-migration of excess oxygen as described by Shi \etal { [24]}; in that experiment, the resistive switching disappeared at temperatures below 200 K because of the low diffusion coefficient of the oxygen. To verify this, the low temperature switching properties were examined using the same structure with a Au electrode. The initial insulating state of this structure is shown in the inset $i$ of Fig. 2(a). After cooling to 4.1 K, the $I-V$ measurements were performed and the structure was negatively biased to $-8.8$ V, at which it switched to a linear state with a very small resistance of $\sim 12$ $\Omega$ as shown in Fig. 2(a), which corresponds to a metallic phase as revealed in the inset $n$ of Fig. 2(a). The voltage required to obtain this switching is larger than the one at room temperature. The sample was then warmed up to 298 K, after which it was switched back to a non-metallic state as indicated by the inset $m$ of Fig. 2(b) with a higher resistance of $\sim 10$ k$\Omega$. These results are consistent with the results seen in the room-temperature device; it is thus very difficult to attribute the reversible MIT observed in LAO to the oxygen vacancy or excess oxygen scenarios.

\begin{figure*}
\includegraphics[width=5.4in]{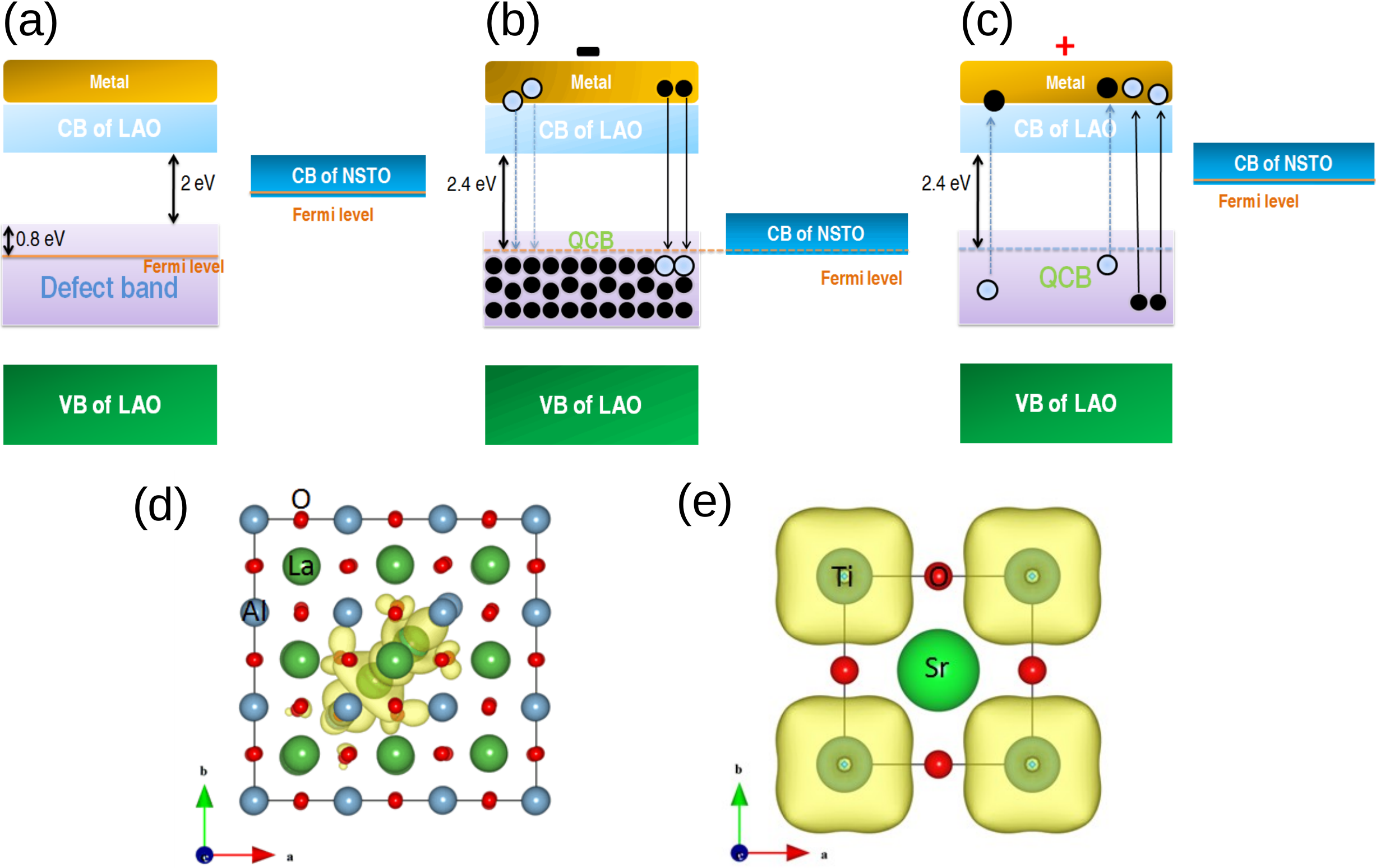}
\caption{\label{fig3}(a) Schematic of the band diagram of the device with no voltage bias. The middle defect band represents the defect levels of LAO, which are widely distributed in the bandgap at $\sim$2 eV below the conduction band. (b) Formation of a QCB under an initial negative voltage bias. (c) Depletion of electrons in the QCB by a subsequent positive bias. (d) Partial charge density distributions of local defect state around the Fermi level for LAO. (e) The lowest unoccupied conduction band at $\Gamma$ point for cubic SrTiO$_3$, respectively. The yellow clouds represent charge densities exceeding 0.011872 electrons/{\AA}$^3$.}
\end{figure*}

We propose a novel mechanism to describe this unusual insulating behavior. Our femtosecond pump-probe experiments on single crystal LAO reveal that defect states can exist in LAO within a wide energy range at approximately 2.0 eV below the conduction band, which is consistent with the theoretical calculations of Luo \etal [25]. In Fig. 3(a), a sketch of the band structure of the device is depicted. For LAO, it is very difficult to identify that it has a $p$-type or $n$-type band structure, so we use the term "Fermi level" just to mark the middle of its bandgap.  Additionally, the work function of the utilized NSTO substrate is very small, determined to be $\sim 3.9$ eV. Therefore, it is plausible to assume that the Fermi level of NSTO is higher than the Fermi level of LAO, as seen in Fig. 3(a). Initially, there is no conduction between the NSTO and LAO. Under sufficient negative bias, charges will be injected into and fill up the LAO defect levels from the metal side and the energy band of NSTO will be pushed down relative to the bands of LAO. The Au/LAO and the NSTO bottom electrode constitute a capacitor. The charges going into the trap states is determined by the applied field. At a critical field where the charge density in the defect states reaches the Mott limit ($\sim3.6\times10^{19}$ cm$^{-3}$), a quasi continuum state is formed where we have assumed an effective electron mass of 1 and a dielectric constant of 20 for the LAO film. We call this defect-mediated continuum state a ``quasi-conduction band" (QCB). Hence one expects for all devices a similar switch on field. The applied voltage will then scale linearly with thickness under the condition that the dielectric constant remains field invariant. As demonstrated in Fig. 3(b), once the energy level of QCB lines up with the Fermi level of NSTO, which is slightly above its conduction band [26],  a continuum state is established, and this state remains even when the current is reduced to zero because of the overlap of the electron wave functions. The effective charge transferred to the defect levels at the switching threshold of 0.22 A is estimated to be around $4.6\times 10^{19}$ cm$^{-3}$ for a device area of 0.02 cm$^2$, a LAO thickness of 150 nm and a defect level lifetime of $\sim 10$ microseconds. While this carrier concentration is about 3.5 times smaller than that of the NSTO, which is $1.6\times 10^{20}$ cm$^{-3}$, it is only slightly larger than the estimated Mott limit, which makes the LAO a stable metallic system. To validate this idea, theoretical calculations on interstitial La$^{2+}$ defect of LAO and percolation of wave functions were performed (supporting material). It was found that the charge density higher than 0.011872 ${\AA}^{-3}$ exists mainly at La atoms around the defect with strong $d$-orbital character, which is of the same order of magnitude as our estimated number. Interestingly, the charge density of the lowest unoccupied conduction band at $\Gamma$ point around Ti atoms in cubic SrTiO$_3$ has the same magnitude as that of the defect state of LAO such that the defect La$-d$ and Ti$-d$ wave functions can couple if they are in proximity, as shown in Fig. 3(d). The most interesting aspect of the QCB is the inherent hysteresis; the only way that this device can be restored to the original insulating state is by applying an opposing voltage, which will then remove the carriers from the QCB and relatively pull the conduction band of NSTO up, leading to the lack of overlap between the QCB and the conduction band of NSTO, as depicted in Fig. 3(c). In the forward regime the LAO and NSTO are indeed metallic and that is the reason why the QCB continues to exist even when the voltage drops below the switch on voltage and even after the polarity is reversed. However, when the voltage reaches the difference between the QCB and the CB of LAO the electrons now can be extracted out of the QCB into the CB of LAO and this causes the QCB to vanish, resulting in the Switch Off. Intriguingly, the transition where the system becomes an insulator (with an applied positive voltage) is close to 2.4 V, which is the difference in the bandgaps of LAO and NSTO. One would predict the switch off voltage to be independent of the LAO thickness.

\begin{figure}
\includegraphics[width=3.4in]{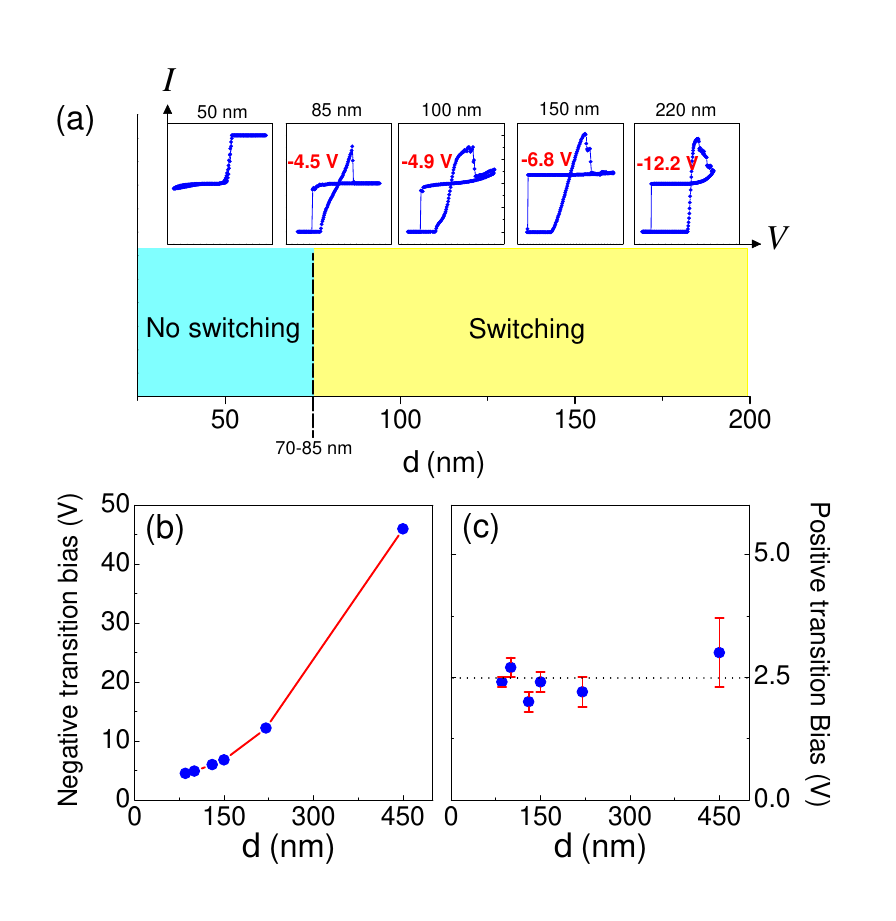}
\caption{\label{fig4}(a) Thickness dependence of the $I-V$ curves for 10, 25, 50, 70, 85, 100, 130, 150, 220 and 450 nm thick LaAlO$_3$ showing a minimum thickness threshold ($70-85$nm) for observing hysteretic switching (only $I-V$'s for 50, 85, 100, 150 and 220 nm shown). (b) The forward switch-off and (c) the reverse switch-on voltage thresholds plotted as a function of LAO thickness for films above 70 nm which exhibit the reversible MIT.}
\end{figure}

This QCB originates from the defect levels in the bandgap of LAO, which is supported by the change in resistive state values after several cycles; the device has slightly lower resistance values after the first cycle, and those values slightly increase in the later cycles as shown in Fig. 1(c) and (d). This change points to the possibility of defect production and defect rearrangement after large currents have passed through these devices. To further test this hypothesis, devices as a function of different LAO thickness were studied. Figure 4a shows the switching behavior as a function of thickness with the $I-V$ characteristics shown for a few of the many devices studied. It is clear that a minimum thickness ($70 - 85$ nm) is required before switching occurs. This is consistent with the defect picture as the films become more defective with increasing thickness. This may arise due to the inherent lattice mismatch between LAO and SrTiO$_3$. In Fig. 4b, the reverse switch-on voltage shows a superlinear dependence on the thickness and this can be argued due to two effects; the need to create a threshold quantity of charge to create the QCB and to account for variations in dielectric constant and also the barrier potential between the metal and LAO as a function of thickness and field. The forward switch-off  threshold voltage is plotted in Fig. 4c and is found to be thickness independent at a value of  $\sim2.4$ V, close to the bandgap difference between LAO  and SrTiO$_3$. The forward switch-off is very sharp for smaller thicknesses (85 nm film) and tends to broaden with increasing thickness (as indicated by error bars) suggesting creation of more defect levels with increasing thickness.   The general trend of the thickness dependence of the threshold voltages in the two polarities support the basis of the proposed model.Further detailed experiments would be needed to fully understand this behavior.

Most of the complex oxides have predominantly ionic bonds and are prone to a variety of cationic and anionic defects including vacancies, interstitials and antisites. These defects create a plethora of electronic states within the bandgap of these oxides. In these insulators, the defect levels can be populated to form QCB, which can lead to multiple conduction states in the same system. These states can be stabilized if an adjacent metallic conduction band overlaps with them. The only way to restore the levels back to their original state is by removing the carriers from the defect levels; this phenomenon is what leads to hysteresis in the $I-V$ curves.

In summary, we have studied the resistive switching of LAO films in metal/LAO/NSTO heterostructures and observed the electric-field-induced reversible MIT. The reversible MIT is ascribed to the population and depletion of QCB consisting of a wide range of defects states in LAO. The stable metallic state can be obtained only when the filling level of QCB inside the LAO aligns with the Fermi level of NSTO such that the wave functions of electrons inside the QCB and the conduction band of NSTO can overlap and interact with each other. The implications of this mechanism are far-reaching especially now the entire semiconductor industry is moving toward high$-k$ materials. For example, the use of multi-component oxides as insulators in devices ({\it{e.g.}}, high$-k$ dielectrics in silicon CMOS devices) must be exercised with caution because of the presence of multiple defect levels within their bandgap.

\begin{acknowledgments}
We thank M. Huijben, J. Mannhart and M. Yang for discussion and experimental support and the National Research Foundation (NRF) Singapore under the Competitive Research Program (CRP) 'Tailoring Oxide Electronics by Atomic Control' NRF2008NRF-CRP002-024, National University of Singapore (NUS) cross-faculty grant and FRC for financial support.
\end{acknowledgments}


\end{document}